\documentclass[fleqn,a4paper]{vch-book}
\usepackage{amsmath}
\usepackage{amssymb}
\usepackage{makeidx}
    \makeindex
\usepackage{times}
\usepackage{tabularx}
\usepackage{booktabs}
\usepackage[dvips]{epsfig}
\DeclareGraphicsExtensions{.eps,.ps}

\chardef\bslash=`\\ 

\newcommand{\bibtex}{\ifx\is@itshape\f@shape{\fontshape{scit}\selectfont
Bib}\else\textsc{Bib}\fi\kern-.1em\TeX}

\newcommand{\one}{\mbox{$1 \hspace{-1.0mm}  {\bf l}$}}  
\newcommand{\ket}[1]{ | \, #1  \rangle}                  
\newcommand{\bra}[1]{ \langle #1 \,  |}                  
\newcommand{\braket}[2]{\left< #1 \right| #2 \rangle}    

\begin{document}

\chapter*{Holonomic Quantum Computation \tocauthor{Angelo C.M. Carollo and Vlatko Vedral}}\label{chap:holonomicComp}

\authorafterheading{Angelo C.M. Carollo$^1$ and Vlatko Vedral$^2$}

\affil{ $^1$Deparment of Applied Mathematics and Theoretical Physics,
Cambridge University, UK\\
 $^2$School of Physics and Astronomy,
University of Leeds,
UK}
A considerable understanding of the formal description of quantum
mechanics has been achieved after Berry's discovery~\cite{Berry84}
of a geometric feature related to the motion of a quantum system.
He showed that the wave function of a quantum object retains a
memory of its evolution in its complex phase argument, which,
apart from the usual dynamical contribution, only depends on the
``geometry'' of the path traversed by the system. Known as the
\emph{geometric phase} factor, this contribution originates from
the very heart of the structure of quantum mechanics.

A renewed interest in geometric phenomena in quantum physics has been
recently motivated by the proposal of using geometric phases for
quantum computation. Geometric
(or `Berry') phases depend only on the geometry of the
path executed, and are therefore resilient to certain types of
errors. The idea is to exploit this inherent
robustness provided by the topological properties of some quantum
systems as a means of constructing built-in fault tolerant quantum
logic gates. Various strategies have been proposed to reach this
goal, some of them making use of purely geometric evolutions, i.e.
non-Abelian holonomies~\cite{ZanardiR99,PachosZR00,PachosC00}.
Others make use of hybrid strategies that combine together
geometrical and dynamical
evolutions~\cite{EkertEHIJOV00,JonesVEC99}, and others yet use more
topological structures to design quantum
memories~\cite{DennisKLP02,Kitaev03}. Several proposals for
geometric quantum computations have been suggested and realised in
different contexts, in NMR experiments~\cite{JonesVEC99}, ion
traps~\cite{DuanCZ01,LeibfriedDMLBBIJLRW03,SorensenM03,Garcia-RipollZC03,StaanumDM04},
cavity QED experiments~\cite{RecatiCZCZ02}, atomic
ensembles~\cite{UnanyanSB99,LiZZS04}, Josephson junction
devices~\cite{FalciFPSV00}, anyonic systems~\cite{Kitaev03},
quantum dots~\cite{SolinasZZR03}.

\section*{Geometric phase and holonomy}

Suppose that a system undergoing a cyclic evolution is described
by classical mechanics; it is impossible to tell from its initial
and final state whether it has undergone any physical motion. The
situation in quantum mechanics is quite different. The state
vector of a quantum system retains the ``history'' of its
evolution in the form of a geometric phase factor. 

This deep and fundamental concept was originally discovered by Pancharatnam~\cite{Pancharatnam56} in the
context of a classical beam of  polarised light and "rediscovered" in a quantum mechanical context by Berry~\cite{Berry84}. Pancharatnam introduced the concept of
{\em parallelism} between two states, as a criterion
to compare the relative phase between two beams of light with different
polarisation. He recognised that a natural convention to measure the phase
difference between two interfering beams is to choose a reference where the intensity has its maximum. For example, by
superimposing two beams of polarisations $\psi_1$ and
$\psi_2$ the intensity is proportional to 
$I\propto
1+|\bra{\psi_1}\psi_2\rangle|\cos\big(\chi+
\arg{\bra{\psi_1}\psi_2\rangle}\big).$
\noindent The interference
fringes are shifted by $\varphi=\arg{\bra{\psi_1}\psi_2\rangle}$, which,
following Pancharatnam's prescription, represents the phase difference between $\psi_1$ and $\psi_2$. 
This idea, translated into
quantum mechanics, leads to the
definition of relative phase between any (non-orthogonal) states lying in a (finite or infinite) Hilbert space. When $\arg{\bra{\psi_1}\psi_2\rangle}=0$, $\psi_1$ and $\psi_2$ are called \emph{in phase}.
Pancharatnam's most important contribution was to point out
that this condition is \emph{not transitive}: if $\psi_1$ is in phase with $\psi_2$ and $\psi_2$ with $\psi_3$, the phase between $\psi_1$ and
$\psi_3$ is, in general, not zero. As in quantum mechanics states are defined up to a phase, $\psi_2$ can always be redefined parallel to $\psi_1$.
However, when a third state
$\psi_3$ is considered, it is, in general, impossible to redefine it in phase with both $\psi_1$ and $\psi_2$. This is due to an {\em irreducible} phase contribution $\chi=\arg \braket{\psi_1}{\psi_2}\braket{\psi_2}{\psi_3}\braket{\psi_3}{\psi_1}$, called Pancharatnam phase\emph{Pancharatnam phase}, 
which represents the most elementary example of \emph{geometric phase}~\cite{Bargmann64,MukundaS93,RabeiMS99}.

If, instead of a discrete collection, we consider a continuous chain of states $\ket{\phi(s)}$ (with $s\in\{s_0\dots s_1\}$), we can repeat a similar argument and redefine the local phases $\ket{\phi(s)}\to\ket{\psi(s)}=e^{i \alpha(s)}\ket{\phi(s)}$ to impose the phase condition between infinitely neighboring states, namely
\begin{equation}\label{paralleltransp}
\arg{\braket{\psi(s)}{\psi(s+ds)}}\simeq \braket{\psi(s)}{\frac{d}{ds}|\psi(s)}ds=0,
\end{equation}
which is known as \emph{parallel transport condition}\index{parallel transport}. As emphasized earlier, this condition is not transitive. Therefore, although neighboring states are in phase, states far apart along the curve accumulates a finite phase difference between them. In particular, if the chain is a closed loop, i.e. $\ket{\phi(s_0)}=\ket{\phi(s_1)}$, a state ``parallel-transported'' around the loop experiences a phase shift 
\begin{equation}
\label{geometricphase}
	\ket{\psi(s_1)}=e^{i\chi_\gamma}\ket{\psi(s_0)}\text{, }\quad \chi_{\gamma}\!=\alpha(\!s_1\!)\!-\!\alpha(\!s_2\!)=i\!\!\!\int_{s_1}^{s_2} \!\!\!\braket{\phi}{\frac{d}{ds}|\phi}ds=i\!\oint_\gamma \braket{\phi }{d\phi}\text{,}
\end{equation} 
which is the celebrated \emph{geometric phase}\index{geometric phase}. As for the Pancharatnam phase, $\chi_{\gamma}$ is an \emph{irreducible} phase contribution which solely depends on the closed path $\gamma$ traced out by $\ket{\psi(s)}$ in the Hibert space. It is easy to verify that neither a \emph{local redefinition of phase}, nor a change in the \emph{rate of traversal} affects the value $\chi_{\gamma}$. 
 
\subsection*{Adiabatic implementation of holonomies}
A natural question to ask is how the idea of parallel transport applies to physical scenarios. It turns out that this concept plays a key role in a variety of physical contexts~(see~\cite{ShapereW89,BohmMKNZ03,Nakahara90}), and, in quantum mechanics it emerges as a natural feature of adiabatically evolving systems. 

Suppose that an Hamiltonian, $H(\lambda_t)$ is controlled by a set of time-dependent parameters $\lambda_t$. If the requirements for the adiabatic approximation (see~\cite{Messiah,Kato50}) are satisfied,  a state, initially prepared in an eigenstate $\ket{\psi(t_0)}=\ket{\Psi_n(\lambda_{t_0})}$, remains eigenstate of the instantaneous Hamiltonian, during the evolution:\index{Adiabatic approximation} 
\begin{equation}
\label{eq:evolvingstate}
\ket{\psi(t)}=e^{i\delta(t)} \ket{\Psi_n(\lambda_{t})} \quad \text{ where } \quad H(\lambda_t)\ket{\Psi_n(\lambda_t)}=\epsilon_n(\lambda_t)\ket{\Psi_n(\lambda_t)},
\end{equation}
($\hbar=0$) where $\delta(t)=-\int_{t_0}^t\epsilon_n(\lambda_t) dt$ is the usual \emph{dynamical phase}.
Under this approximation, the state $\ket{\psi(t)}$ can satisfy the Schr\"odinger equation only if the contraint  $\braket{\Psi_n(\lambda_t)}{\frac{d}{dt}|\Psi_n(\lambda_{t})}=0$ is fufilled.
Hence, the state $\ket{\Psi_n(\lambda_t)}$ is parallel transported around the Hilbert space as the parameters $\lambda$'s are varied. If the latter are eventually brought back to their initial values $\lambda_0$, and the eigenspace of $\ket{\Psi_n}$ is \emph{non-degenerate}, the final state will be proportional to the initial one, $\ket{\Psi(t_f)}=e^{i\chi_\gamma}\ket{\Psi(t_i)}$, with an accumulated geometric phase $\chi_\gamma$ (which in this context is called \emph{Berry phase}\index{Berry phase}), only dependent on the path, $\gamma$, traced in the parameter space: 
\begin{equation}
\label{eq:connection}
\chi_\gamma=\oint_\gamma A_i d\lambda_i, \qquad A_i=i\braket{\Psi_n}{\partial_{\lambda_i}\Psi_n}
\end{equation}
where the path integral here is explicitly expressed in terms of a vector (one-form), known as \emph{Berry connection}. 
The inherently geometric nature is even more evident when the path integral in eq.~(\ref{eq:connection}) is formulated as a surface integral, via the Stokes theorem:
\begin{equation}
\chi_{\gamma}=\oint_\gamma \braket{\phi }{d\phi}=\int\!\!\!\int_{\Sigma} \mathcal{F}d\sigma
\end{equation}
where $\Sigma$ is the surface enclosed within the loop traced by $\lambda$ in the parameters' manifold, and $\mathcal{F}_{ij}=\partial_i A_j-\partial_j A_i$ is called the Berry curvature. The Berry curvature in many interesting cases (such as for \emph{qubits}) is a slowly varying function, or even a constant. As a result of this, the geometric phase behaves as an \emph{area} and depends almost exclusively on the \emph{surface} enclosed by the loop. This is one of the crucial characteristic that makes the geometric phase quite appealing for the implementations of \emph{fault-tolerant} quantum computation. A feature, like an area, which is much less dependent on the details of the time evolution, is likely to be less affected by variations of environmental conditions, and hence, \emph{more robust}.

The prototypical example in which this \emph{area-like} behaviour is manifest, is the case of a single qubit adiabatically evolving under a generic Hamiltonian $H(t)=\vec{\mathbf{n}}_t\cdot \vec{\sigma}$, where $\vec{\sigma}=(\sigma_x,\sigma_y,\sigma_z)$ is the vector of Pauli matrices, and $\vec{\mathbf{n}}_t$ is a time-dependent vector.
It is possible to show that the curvature associated with a qubit state gives rise to a very simple form of the geometric phase, namely $\chi_{\gamma}=\pm\frac{\Omega}{2}$
 ($\pm$ depending on whether the qubit is initially aligned or against the direction of $\vec{\mathbf{n}}$), where $\Omega$ is the solid angle spanned by the direction of the vector $\vec{\mathbf{n}}$. The curvature in this case is constant $(\pm 1/2)$ and $\Omega$ is the surface enclosed in parameter manifold (the Bloch sphere)~( see Fig.~\ref{figure}a).

Before turning the discussion towards the implementation of quantum computation, it is important to introduce the \emph{non-Abelian} generalisation of the geometric phase, or \emph{holonomy}\index{holonomy}.
In obtaining the geometric phase for an adiabatic evolving system, the assumption that the eigenspace to which the prepared state belongs is non-degenerate was crucial. Such a condition insures that, when a loop in the parameter space is traversed, final and initial states are proportional: i.e. the net effect of the evolution is merely a phase. However, assuming a degenerate eigenspace, opens up a wider variety of possible evolutions, with a slightly more complex structure, known formally as \emph{holonomy}. 

The word holonomy refers to the set of all the closed curves, or \emph{loops} on a manifold, starting and ending in the same point $x_0$. It is easy to verify that this set has the structure of \emph{group}\footnote{The composition of two loops and is obtained by joining the end point of one loop with the starting point of the other. The identity element is the trivial loop with only one point ($x_0$). The inverse of curve is the same traversed in the opposite direction. For a rigorous definition see~\cite{Nakahara90,Frankel00})}. 
The geometric phases themselves form a \emph{representation} of an holonomy group: any loop in the parameter space of an Hamiltonian is associated with a geometric phase factor. And clearly they form an \emph{Abelian representation} as phases commute: $e^{i\chi_{\gamma_1}}e^{i\chi_{\gamma_2}}=e^{i\chi_{\gamma_2}}e^{i\chi_{\gamma_1}}$. 
This therefore implies that their non-Abelian
generalisation are not represented by ordinary numbers, but by
matrices. This naturaly emerges in adiabatic evolving systems, when eigenspaces are degenerate.\\
Let's write a parameter dependent Hamiltonian in the form: $H(\lambda_t)  =  \sum_k \epsilon_k(\lambda_t) \Pi_k(\lambda_t)$, where $\Pi_k(\lambda_t)$ are the projector operators of the instantaneous eigenspaces. As time varies, the parameters change and with them eigenvalues and eigenspaces. The latter are smoothly concatenated via a unitary transformation $O(\lambda_t)$ (the eigenspaces never change dimension, as this is forbidden by the adiabatic requirements), $\Pi_k(\lambda_t)=O(\lambda_t)\Pi_k^0O^\dag(\lambda_t)$, where $\Pi_k^0$ is an eigenspace at the initial time $t_0$ ($O(\lambda_{t_0})=\one$). The unitary transformation $O^\dag$ produces the change of picture to the frame moving rigidly with the instantaneous eigenspaces. In this frame, the evolution is governed by the Hamiltonian 
$\tilde{H}(\lambda_t)=\sum_k \epsilon_k \Pi_k^0-i\frac{dO(\lambda_t)}{dt}O^\dag(\lambda_t)$.
Imposing the adiabatic approximation\index{Adiabatic approximation} is equivalent to neglecting Hamiltonian terms coupling different eigenspaces~(see~\cite{Messiah}). The evolution inside each eigenspace is, then, generated by the following equation:
\begin{equation}
\label{scrhodinger}
i\frac{dU_k(t)}{dt}=\left[\epsilon(\lambda_t)-A_k(\lambda_t)\right] U_k(t), \qquad A_k(\lambda_t)=i\Pi_k^0\frac{dO(\lambda_t)}{dt}O^\dag(\lambda_t)\Pi_k^0\text{.}
\end{equation}
This equation can be formally solved, and, for a closed loop of the parameters, yields the total evolution (notice that by definition $O(\lambda_{t_f})=\one$): 
\begin{equation}
\label{ }
U_k(t_f)=T_k(t_f)V_k^\gamma, \qquad \text{with } \quad V_k^{\gamma}=\mathcal{P}\exp\oint_{\gamma} A_k(\lambda)d\lambda, \text{ and} \quad T_k=e^{-i\int \epsilon_k t}\text{,}
\end{equation}
where $T$ is an overall dynamical phase factor, and $V_k^{\gamma}$ is the celebrated (\emph{non-Abelian}) \emph{holonomy}. In this formula $\mathcal{P}$ is the \emph{path}-ordering operator, needed because of the non-commutativity of the operators $A(\lambda)$ 
for different values of the parameters. 
This non-Abelian phases is in
general very difficult to evaluate, because of the path ordering
operation. 

\section*{Application to quantum computation}
We would like to mention potential advantages of using geometrical
evolution to implement quantum gates. First of all, there is no
dynamical phase in the evolution. This is because we are using
degenerate states to encode information so that the dynamical
phase is the same for both states (and it factors out as it were).
Also, all the errors stemming from the dynamical phase are
automatically eliminated. Secondly, the states being degenerate do
not suffer from any bit flip errors between the states (like the
spontaneous emission). So, the evolution is protected against
these errors as well. Thirdly, the size of the error depends on
the area covered and is therefore immune to random noise (at least
in the first order) in the driving of the evolution. This is
because the area is preserved under such a noise as formally
proven by DeChiara end Palma~\cite{DeChiaraP03}. Also, by tuning the parameters of
the driving field it may be possible to make the phase independent
of the area to a large extent and make it dependent only on a
singular topological feature - such as in the Aharonov-Bohm effect
where the flux can be confined to a small area - and this would
then make the phase resistant under very general errors.

So, in order to see how this works in practice we take an atomic
system as our model implementing the non-Abelian evolution. We'll
see that quantum computation can easily be implemented in this
way. The question, of course, is the one about the ultimate
benefits of this implementation. Although there are some obvious
benefits, as listed above, there are also some serious
shortcomings, and so the jury is still out on this issue.

\subsection*{Example}
Let's look at the following $4$ level system analyzed by Unanyan,
Shore and Bergmann~\cite{UnanyanSB99}. They considered a four level
system with three degenerate levels $1,3,4$ and one level $2$ with
a different energy as in Fig~\ref{figure}b. This system stores one bit of
information in the levels $1$ and $2$ (hence there is double the
redundancy in the encoding of information). We have the following
Hamiltonian
$$
H(t) = \left(\begin{array}{cccc} 0 & P(t) & 0 & 0 \\
P(t) & 0 & S(t) & Q(t) \\ 0 & S(t) & 0 & 0 \\ 0 & Q(t) & 0 & 0
\end{array}\right)
$$
where $P,Q,S$ are arbitrary functions of time. It is not difficult to find eigenvalues and eigenvectors
of this matrix (exercise!). There are two degenerate eigenvectors
(with the corresponding zero eigenvalue for all times) which will
be implementing our qubit and they are
\[
\Phi_1 (t) = \left( \cos \theta_t,0,-\sin \theta_t 0 \right)\quad\text{ and }
\quad 
\Phi_2 (t) = \left(\sin \phi_t \sin\theta_t,
0,\sin\phi_t\cos\theta_t,-\cos \phi_t \right)
\]
where $\tan \theta_t = P(t)/Q(t)$ and $\tan \phi_t =
Q(t)/\sqrt{P(t)^2+Q(t)^2}$.
\begin{vchfigure}[t]\label{figure}
\includegraphics[width=0.7\textwidth]{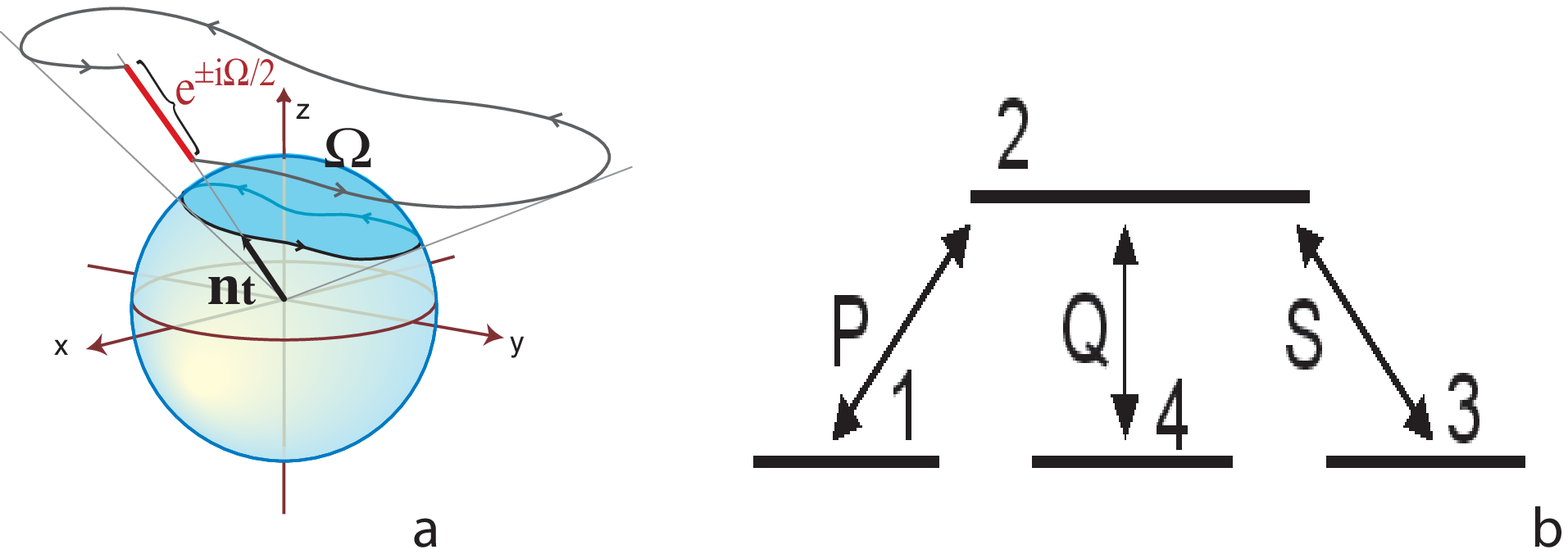}
\vchcaption{(a) The geometric phase for a single qubit is proportional to the solid angle $\Omega$. (b)The four level system that can be used for non-Abelian quantum
computation to encode one qubit of information in two degenerate
levels. The method is detailed in the text.}
\end{vchfigure}
In the adiabatic limit, we can restrict ourself to these states only. Although, in general, the Dyson equation is difficult to solve, in this special example we can write down a closed form
expression~\cite{UnanyanSB99}. The unitary matrix representing the
geometrical evolution of the degenerate states is
\begin{equation}\label{matrixB}
B(\eta_t) = \left(\begin{array}{cc} \cos \eta_t & \sin \eta_t
\\ -\sin \eta_t & \cos \eta_t \end{array}\right)
\end{equation}
where $\eta_t = \int_0^t \sin \phi_\tau \frac{d\theta}{d\tau} d\tau$.
This therefore allows us to calculate the non-Abelian phase for
any closed path in the parametric space. After some time we
suppose that the parameters return to their original value. So, at
the end of the interaction we have the matrix $B(\eta_f)$
where $\eta_f = \oint_c \frac{Q}{(P^2 + S^2)\sqrt{Q^2 + P^2 + S^2}}
(SdP - PdS)$,
which can be evaluated using Stokes' theorem (the phase will in
general depend on the path, as explained before). So, we can have
a non-Abelian phase implementing a Hadamard gate. With two systems
of this type (mutually interacting) we can implement a controlled-Not gate 
and therefore (at least in principle) have a universal
quantum computer~(see \cite{ZanardiR99}). 

\renewcommand\bibname{References}

\end{document}